# Adaptive Dual Channel Convolution Hypergraph Representation Learning for Technological Intellectual Property

Yuxin Liu[1], Yawen Li[2*], Yingxia Shao[1], Zeli Guan[1]

[1]Beijing Key Laboratory of Intelligent Communication Software and Multimedia, School of Computer Science (National Pilot Software Engineering School), Beijing University of Posts and Telecommunications, Beijing 100876
[2]School of Economics and Management, Beijing University of Posts and Telecommunications, Beijing 100876

**Abstract:** In the age of big data, the demand for hidden information mining in technological intellectual property is increasing in discrete countries. Definitely, a considerable number of graph learning algorithms for technological intellectual property have been proposed. The goal is to model the technological intellectual property entities and their relationships through the graph structure and use the neural network algorithm to extract the hidden structure information in the graph. However, most of the existing graph learning algorithms merely focus on the information mining of binary relations in technological intellectual property, ignoring the higher-order information hidden in non-binary relations. Therefore, a hypergraph neural network model based on dual channel convolution is proposed. For the hypergraph constructed from technological intellectual property data, the hypergraph channel and the line expanded graph channel of the hypergraph are used to learn the hypergraph, and the attention mechanism is introduced to adaptively fuse the output representations of the two channels. The proposed model outperforms the existing approaches on a variety of datasets.

## 1 Introduction

Technological intellectual property data is generally composed of complex information like creative entities, resource entities, inter-entity relationships, and entity attributes[33]. Mainstream models usually need to map data into a vector space that is convenient for calculation[20][23],[27], filter and fuse data in the limited space[39]-[43], and cannot ignore the correlation information of the data itself. The purpose of this paper is to propose a hypergraph learning algorithm to represent the hypergraph modeled the complex information about technological intellectual property data for further analysis and mining.

In terms of the modelling technological intellectual property data, existing methods include a homogeneous [30],[32] and a heterogeneous graph [25][26],[31]. However, homogeneous and heterogeneous graphs merely model binary relationships, which are not able to model non-binary relationships in technological intellectual property data. Consequently, graph learning algorithms based on a homogenous graph and a heterogeneous graph are unable to represent nonlinear higher-order correlation among technological intellectual property entities. Some researchers have proposed to use multi-agent to learn multiple features and solve the above problems through feature fusion [12-15], but this method requires a lot of artificial construction work. Instead, hypergraph [29] has been used as a modeling method of non-binary relations due to its structural characteristics, such as the research [10], [46] on natural language processing, the research [12],[34]-[36] on recommendation systems, the research [11],[45] on image classification, and the research on various social networks [38],[44]. Accordingly, using hypergraph to model the entities and relationships of technological intellectual property data is able to fully express its complexity.

For hypergraph learning, traditional algorithms are typically separated into two categories: one is based on incidence matrix decomposition, and the other is based on hypergraph expansion [28],[37], which converts hypergraphs into weighted simple graphs. Neither the performance nor the efficiency of traditional methods can be compared with neural networks, so this paper mainly discusses hypergraph neural network algorithms. The neural network algorithms include hypergraph neural networks based on hypergraph expansion and hypergraph neural networks based on non-hypergraph expansion. Methods [1][3] based on hypergraph expansion commonly use simple graph learning methods such as the Graph Convolution Network (GCN) to represent the converted simple weighted graph from the hypergraph. However, it would cause the loss of some information for the same reason that it simplifies the hypergraph into a simple weighted graph. The methods [5][7] based on non-hypergraph expansion are to carry out hypergraph learning through hypergraph convolution neural networks or hypergraph attention neural networks. Some of these methods are merely appropriate for learning k-uniform graphs with k hypernodes on each hyperedge but not for the technological intellectual property data that hypergraph structure is a k-uniform graph. The others [8] integrate the features of hypergraph and simple graph expanded from hypergraph. Nevertheless, the method of fusion is the average of the representation of two graphs.

In view of the problems with the above methods, we propose an adaptive dual-channel hypergraph node representation algorithm based on hypergraph convolution and line expansion (LE) of the hypergraph. Furthermore, it uses the node classification task to verify the method. Firstly, we construct a hypergraph based on some technological intellectual property data by taking the author as the hyperedge and his publications as the hypernodes. Secondly, using the method proposed by [3] expands the hypergraph to a weighted simple graph. Thirdly, the hypergraph and the graph of LE pass through the hypergraph convolution channel and the graph convolution channel, respectively. Finally, the final node representation is obtained by using the attention mechanism to adaptively fuse the two representations of dual channels. Furthermore, we use the node classification task to determine the effect of node representation.

*Corresponding author: Yawen Li (warmly0716@126.com).

The main contributions of this paper are as follows:

1) We propose an adaptive hypergraph dual channel convolution node representation learning method (ADHCN). The hypergraph line expansion convolution channel and the hypergraph convolution channel are used to represent nodes, so as to avoid the information loss in the simple graph, which is line expanded from the hypergraph.

2) We introduce the attention mechanism to fuse the representations of the two channels to avoid the deviation caused by direct averaging or splicing.

3) The experimental results demonstrate that ADHCN exceeds the compared models in terms of accuracy, F1, and recall. Moreover, the experiment shows that the effect of the model is significantly improved after introducing the attention mechanism for adaptive fusion.

## 2 Related Work

A hypergraph $G = <V, E, W>$ is a generalized graph, where $V = \{v_1, v_2, ..., v_n\}$ is the set of hypernodes, $E = \{e_1, e_2, ..., e_m\}$ is the set of hyperedges, and $W = \{w_1, w_2, ..., w_m\}$ is the weights set on each hyperedge. In particular, each hyperedge $e_i$ connects more than two hypernodes. Moreover, the incidence matrix of the hypergraph is $H = \{h_{ij} | 1 \leq i \leq n, 1 \leq j \leq m\}$, where $h_{ij}$ only takes 0 or 1. In particular, there are more than two 1 in the column of H, which means one hyperedge connects several vertices.

For the methods in hypergraph neural networks based on expansion, Yadati et al. [1] proposed the hyperGCN model, whose principle is to use the GCN training graph transformed from the hypergraph through the spectral theory and introduce the mediator to prevent the loss of the hypergraph's information. Bandyopadhyay [2] simplified the hypergraph into a normal graph named "line graph" by taking the hyperedge as the vertex and the Jaccard similarity between the hyperedges as the weight, and then trained the graph with GCN. Yang et al. [3] proposed the line expansion (LE) of hypergraphs to transform the hypergraph into a weighted simple graph $G_l = (V_l, E_l)$. The $V_l$ is the node set which is composed of hyperedge-hypernode pairs $\{(v, e) | v \in V, e \in E\}$ from the original hypergraph. The edge of the edge set $E_l$ connects two pairs if both pairs have a common hypernode or hyperedge. The adjacency matrix of the weighted graph $A_l \in \{0, w_e, w_v\}^{|V_l| \times |V_l|}$ shown in Eq.1 is defined by the pairwise relation $u \in V_l$ and $v \in V_l$.

$$A_l(u,v) \begin{cases} w_e \ u = (v_h, e_h), v = (v'_h, e'_h), v_h = v'_h \\ w_v \ u = (v_h, e_h), v = (v'_h, e'_h), e_h = e'_h \\ 0 \ otherwise \end{cases} \quad (1)$$

The above algorithms reduce the hypergraph learning problem with simple graph learning, which brings problems of hypergraph information loss. For the methods in hypergraph neural networks based on non-expansion, Feng et al. [5] proposed a hypergraph convolution neural network by extending graph convolution to hypergraph. And the proposed method is used to learn the hypergraph constructed by k-nearest neighbor (KNN). Owing to KNN being the method of constructing hypergraphs, this model is merely appropriate for k-uniform hypergraphs. In order to address the issue that the hypergraph structure in [5] changes as a result of the alteration in node representation. Jiang et al. [6] proposed a dynamic hypergraph neural network (DHGNN) that contains dynamic hypergraph reconstruction that reconstructs the hypergraph at each layer and dynamic graph convolution that gathers the information of nodes and edges. However, the method is incapable of solving the k-uniform graph problem. Bai et al. [7] proposed a hypergraph convolution theory from another perspective, which is consistent with the hypergraph convolution proposed by [5]. Besides, it introduced hypergraph attention to enhance the expression ability. Xia et al. [8] proposed a dual channel hypergraph convolution network to solve the problem of session-based recommendation. In particular, the model uses hypergraph convolution and line graph convolution to learn hypergraph and fuse the learning result. In spite of that, there are still some problems in the model, such as information loss in the line graph channel and the simplicity of the fusion method. Yu et al. [9] proposed a multi-channel hypergraph convolutional network for social recommendation. The model builds three hypergraphs and uses three hypergraph convolutions to represent these graphs. However, the method can't be used in the technological intellectual property dataset.

## 3 ADHCN: Adaptive Dual Channel Hypergraph Convolution Network

Figure 1 shows the model structure proposed in this paper. It mainly includes the hypergraph construction and the line expansion (HCLE) module, the dual channel convolution (DHC) module, the adaptive representation fusion (ARF) module, and the model verification module.

Firstly, the dataset is constructed into a hypergraph H and a line-expanded simple graph $G_l$ through the HCLE module. Moreover, the initial representation of the two graphs is also the output of the module. Secondly, the output of the HCLE module is the input of the DHC module. Thirdly, the ARF module fuses the outputs of the two channels to obtain the final representation. Finally, the representation is input to a classifier to distinguish the hypernode and supervise the model training.

### 3.1 HCLE: Hypergraph Construction and Line Expansion

ADHCN constructs entities and relationships in technological intellectual property data as hypergraphs $G_h = <V, E>$, in which publications are viewed as hypernodes $V$ and authors as hyperedges $E$. The incidence matrix $H \in \{0,1\}^{|V|\times|E|}$ with $H(a, p_a) = 1$ is defined by the relationship between an author $a$ and his publications $P_a = \{p_1, p_2, ..., p_t\}$. The weights of all hyperedges of the hypergraph are set to 1. And the initial representations $X_h$ of the publications are the text embeddings of their abstracts.

ADHCN adopts the line expansion of the hypergraph as the input of another convolution channel. The line expansion method of the hypergraph of technological intellectual property data is to treat the author-publication pair as nodes to obtain a weighted simple graph $G_l = \{<p, a> | p \in V, a \in E\}$. And, there is one edge between the adjacency matrix $A_l$ and incidence matrix H are respectively input to the two channels to train the model.

### 3.2 DHC: Dual Channel Convolution Module

The dual channel convolution module is divided into two channels of convolution, one is line expansion (LE) convolution and the other is hypergraph convolution.

ADHCN adopts the convolution form proposed by GCN as the single-layer calculation method of the LE convolution channel, and in order to reduce the complexity of the model, the number of layers is set to 1. Accordingly, the output of the LE convolution channel is $\widetilde{Z_l}$.

$$\widetilde{Z_l} = \sigma(D^{-\frac{1}{2}}\widetilde{A_l}D^{-\frac{1}{2}}X_l\Theta_l) \qquad (3)$$

where $\sigma(\cdot)$ is the nonlinear activation function, $D^{-\frac{1}{2}} \in R^{|V_l|\times|V_l|}$ is the expanded graph node degree matrix, $\widetilde{A_l} = A_l + 2I \in R^{|V_l|\times|V_l|}$ is the adjacency matrix with the adjustment by adding two-orders of self-loop, $X_l \in R^{|V_l|\times d}$ is the LE channel initial input, $\Theta_l \in R^{d\times h}$ is the channel parameters to be learned.

After the node representation $\widetilde{Z_l}$ of the LE graph is obtained by graph convolution, it needs to be mapped into two pairs if either two authors create the same publication or one author creates two publications. As suggested by [3], the weights $w_e$ and $w_v$ of the adjacency matrix $A_l$ are both 1. In order to obtain the initial input $X_l$ of the line expansion convolution channel, the initial input of the hypergraph convolution channel $X_h$ is converted according to the hypernode projection matrix $P_v$ proposed in [3]. The construction of the hypernode projection matrix is shown in Eq.2.

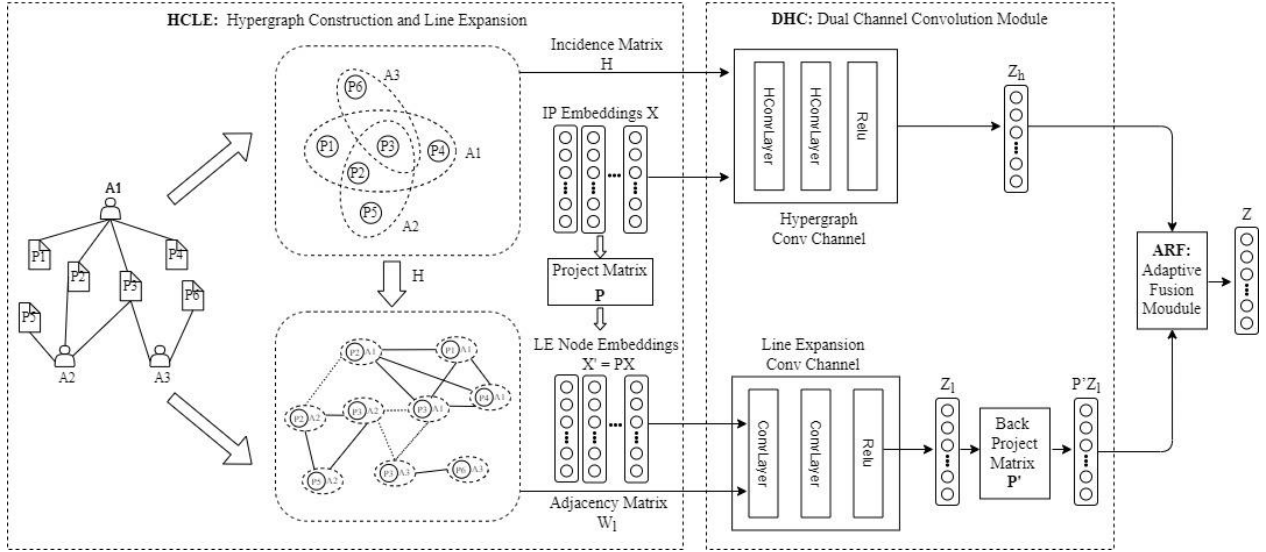

**Figure 1** ADHCN model structure

$$P_v(v_l, v) = \begin{cases} 1 & v_l = (v, e), \exists e \in E \\ 0 & otherwise \end{cases} \qquad (2)$$

Afterwards, the initial inputs $X_l = P_v * X_h$, $X_h$ and the back to the hypergraph to obtain the output $Z_l$ of the LE convolution channel. The construction of hypernode projection matrix is shown in Eq.4.

$$P'_v = \begin{cases} \frac{\frac{1}{\sigma(e)}}{\Sigma\frac{1}{\sigma(e)}} & v_l = (v,e), \exists e \in E \\ 0 & otherwise \end{cases} \qquad (4)$$

where $\sigma(e)$ is the degree of the hyperedge $e$, $E$ is the hyperedge set. Therefore, $Z_l = P'_v * \widetilde{Z_l}$ is the output of the LE convolution channel.

ADHCN adopts the hypergraph convolution form proposed in HGNN [5] as the single-layer calculation rule of the hypergraph convolution channel. In order to prevent overfitting problems caused by too complex models, the proposed hypergraph convolution channel sets the number of layers to 1. Accordingly, the output of the hypergraph convolution channel is $Z_h$.

$$Z_h = \sigma(D_v^{-\frac{1}{2}}HWD_e^{-1}H^T D_v^{-\frac{1}{2}}X_h\Theta_h) \quad (5)$$

where $\sigma(\cdot)$ is the nonlinear activation function, $H \in R^{|V|\times|E|}$ is the hypergraph incidence matrix, $W \in R^{|E|\times|E|}$ is the hyperegde weight matrix, $D_v^{-\frac{1}{2}} \in R^{|V|\times|V|}$ is the hypernode degree matrix, $D_e^{-1} \in R^{|E|\times|E|}$ is the hyperedge degree matrix, $X_h \in R^{|V|\times d}$ is the hyperedge initial input, $\Theta_h \in R^{d\times h}$ is the parameters to be learned. Because the hypergraph weight is set to 1 in this paper, Eq.4 can be simplified to Eq.6.

$$Z_h = \sigma(D_v^{-\frac{1}{2}}HD_e^{-1}H^T D_v^{-\frac{1}{2}}X_h\Theta_h) \quad (6)$$

### 3.3 ARF: Adaptive Representation Fusion Module

The adaptive representation fusion module is mainly used to fuse the node representations learned by the two channels. The main principle is adaptive fusion through attention. Figure 2 shows the structure of this module.

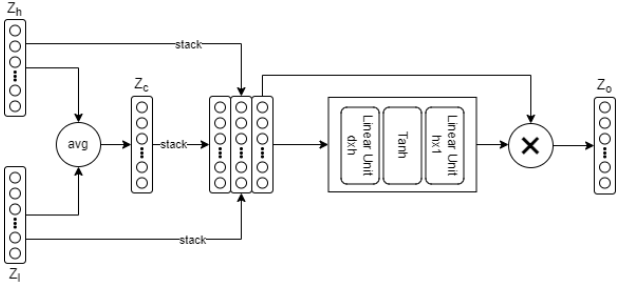

**Figure 2** Adaptive Representation Fusion Module

After obtaining the output representations of the two channels, averaging them to obtain the common representation $Z_c$, then stacking the three representations together and inputting them to the attention module to obtain the weight $W = [\omega_1, \omega_2, \omega_3]$. The weight calculation process is shown in Eq.7.

$$W = [\omega_1, \omega_2, \omega_3] = \text{Attention}([Z_l, Z_h, Z_c]) \quad (7)$$

where $\text{Attention}(\cdot)$ is the attention function and $Z_c = \frac{Z_l+Z_h}{2}$. Finally, use the calculated weight to sum the three representations to obtain the final fusion feature $Z_{out}$.

$$Z_{out} = \omega_1 * Z_l + \omega_2 * Z_h + \omega_3 * Z_c \quad (8)$$

### 3.4 Loss Function

We use hypernode classification as a task to guide model training and verify model performance. The fused features are classified by a linear classifier and softmax to obtain the classification output z.

$$z = \text{softmax}(W_{classifier} * Z_{out} + b_{classifier}) \quad (9)$$

where $W_{classifier} \in R^{h\times c}$ is the weights of the classifier model, $b_{classifier}$ is the bias of the classifier model, z is the classifier output.

The classification loss is computed using the cross entropy loss function, as is typical for most models. The calculation of classification loss is shown in Eq.10.

$$L_c = -\frac{1}{n}\sum[y\ln z + (1-y)\ln(1-z)] \quad (10)$$

where $y$ is the hypernode's label, $z$ is the output of the classifier.

---

**Algorithm 1 ADHCN**

**Input:** Hypergraph incidence matrix $A$, hypernode representations $X_h$
**Input:** Hypergraph-LE project matrix $P_v$, LE-Hypergraph back project matrix $P_v'$
**Input:** Set of target labels $\{y_{true}\}_{k \in L}$
    **for** $i = 1$ to $n\_epochs$ **do**
        **Hypergraph Channel:**
        $Z_h = HConv(H, X_h) = \sigma(D_v^{-\frac{1}{2}}HD_e^{-1}H^T D_v^{-\frac{1}{2}}X_h\Theta_h + b_h)$
        **Hypergraph-LE Projection:** $X_l = P_v X_h$
        **LE Graph Channel:**
        $Z_l = GConv(A, X_l) = \sigma(D^{-\frac{1}{2}}\bar{A}_l D^{-\frac{1}{2}}\Theta_l + b_l)$
        **LE-Hypergraph Back Projection:** $Z_l = P_v' Z_l$
        **Compute Fusion Weights:**
        $[\omega_1, \omega_2, \omega_3] = Attention([Z_l, Z_h, \frac{Z_l+Z_h}{2}])$
        **Fusing Dual Channel:** $Z_{out} = \omega_1 * Z_l + \omega_2 * Z_h + \omega_3 * Z_{c \in L}$
        **Classification:** $y_{pre} = classifier(Z_{out})$
        Compute cross-entropy loss between $\{y_{true}\}_{k \in L}$ and predictions using $\{y_{pre}\}_{k \in L}$
        **Backpropagate** on loss and **optimize** parameters e.g. $\Theta_h$, $\Theta_l$, $b_h$, $b_l$
    **end for**
**return** Learned hypernode representation $Z_{out}$

---

## 4 Experiments

### 4.1 Experimental Settings

We evaluate the quality of ADHCN representation on four datasets through the node classification task. The four datasets are Dblp [17], Citeseer [18], Cora [19], and patentDB. The first three are the commonly used benchmark datasets, while the latter is a self-built dataset in this paper. The self-built dataset named patentDB built by this paper consists of Chinese patents and their applicants and inventors, in which the applicants are the hypernodes of the hypergraph and the patents are the hyperedges of the hypergraph. All patent abstracts owned by the applicants constitute the description of the applicants and are transformed into the representations of the applicants. The applicant categories are companies and individuals.

**Table I** The attributes of different datasets

| Dataset | |V| | |E| | density | classes |
|---|---|---|---|---|
| **Cora** | 12998 | 6250 | 4.145 | 10 |
| **Citeseer** | 1498 | 1107 | 3.178 | 6 |
| **Dblp** | 41302 | 22363 | 4.452 | 6 |
| **PatentDB** | 68275 | 20715 | 3.690 | 2 |

Table I shows the attributes of these datasets, including the number of hypernodes |V|, the number of hyperedges |E|, the density of the hyperedges, and the number of categories.

To optimize the model, we utilize the Adam optimizer. In addition, we chose 0.001 and 0.0005 for the initial learning rate and decay. In terms of model structure, we set the layers of the hypergraph convolution channel and graph convolution channel to 1. Moreover, we set the dropout to 0.5 to avoid overfitting and reduce running time.

For the evaluation indicators, we use the commonly used classification accuracy, macro-F1, and recall. In this study, accuracy, macro-F1, and recall are denoted by the letters Acc, F1, and R, respectively.

In terms of comparison algorithms, we use the following hypergraph learning algorithms:

- **HyperGCN** [1]: In order to reduce the information loss caused by the transformation of a hypergraph into a simple weighted graph and accelerate the training process. Yadati et al. proposed the mediation mechanism, and the accelerated computing mechanism respectively. The existence of these two mechanisms makes the model have multiple forms. No accelerated computation or mediation are employed in 1-HyperGCN. The version of Fast-HyperGCN employs accelerated computing rather than mediation. Mediations are employed in HyperGCN* but not accelerated computing.
- **LE-GCN** [3]: Yang et al. proposed a method named hypergraph line expansion (LE) to transform the hypergraph into a simple weighted graph. In this paper, in order to generate a representation of the hypernodes as a comparison of ADHCN outputs, we utilize GCN to learn the simple graph transferred from a hypergraph via line expansion.
- **HGNN** [5]: Feng et al. proposed a hypergraph construction module based on Euclidean distance algorithm KNN and a hypergraph convolution module based on the hypergraph spectral theory. The hypergraph construction module is not relevant for this article because the hypergraph has already been built. Rather, we directly utilize the hypergraph convolution module to learn the constructed hypergraph.

### 4.2 ADHCN Validation

We utilized the methodology specified in the experimental design for comparison trials to confirm the efficacy of ADHCN. The experimental results are shown in Table II.

The performance of the proposed model in multiple datasets (such as Citeseer, Dblp, and patentDB) is better than that of the comparison algorithm, among which Dblp has the most obvious effect, and the accuracy is improved by 1.4% - 8.4% compared with other models. However, in terms of the Cora dataset, our model is inferior to LE-GCN, especially in recall and F1.

**Table II** Comparison of ADHCN and other models

| Model | Cora | | | Citeseer | | |
|---|---|---|---|---|---|---|
| | Acc | R | F1 | Acc | R | F1 |
| 1-hyperGCN | 0.601 | 0.386 | 0.391 | 0.679 | 0.637 | 0.635 |
| fast-hyperGCN | 0.626 | 0.423 | 0.449 | 0.675 | 0.636 | 0.635 |
| hyperGCN | 0.635 | 0.442 | 0.471 | 0.681 | 0.638 | 0.637 |
| LE-GCN | 0.646 | **0.464** | **0.503** | 0.702 | 0.649 | 0.643 |
| HGNN | 0.639 | 0.451 | 0.494 | 0.680 | 0.639 | 0.634 |
| ADHCN(our) | **0.646** | 0.443 | 0.487 | **0.706** | **0.650** | **0.647** |

| Model | Dblp | | | PatentDB | | |
|---|---|---|---|---|---|---|
| | Acc | R | F1 | Acc | R | F1 |
| 1-hyperGCN | 0.789 | 0.783 | 0.778 | 0.798 | 0.500 | 0.462 |
| fast-hyperGCN | 0.815 | 0.806 | 0.805 | 0.803 | 0.498 | 0.465 |
| hyperGCN | 0.822 | 0.811 | 0.813 | 0.815 | 0.501 | 0.465 |
| LE-GCN | 0.859 | 0.851 | 0.853 | 0.849 | 0.509 | 0.501 |
| HGNN | 0.830 | 0.820 | 0.823 | 0.845 | 0.506 | 0.501 |
| ADHCN(our) | **0.873** | **0.862** | **0.867** | **0.852** | **0.513** | **0.503** |

ADHCN obtains the higher-order information of the expanded hypergraph from the LE channel and adaptively fuses it with the convolution channel representation of the hypergraph. This avoids the loss of some information in the expanded hypergraph and the offset caused by weighted fusion, so it has better performance.

### 4.3 Adaptive Representation Fusion Module Validation

In order to verify the effectiveness of the adaptive fusion module, we compare the performance of ADHCN and the model that obtains the fused feature by $Z_c = Z_h + \alpha Z_l$. Furthermore, the weight ratio $\alpha$ is set to 0.1, 0.3, 0.5, 0.7 and 0.9 respectively. Table III displays the experiment's findings across four datasets.

**Table III** Ablation Experiment on Cora and Citeseer

| Model | Cora | | | Citeseer | | |
|---|---|---|---|---|---|---|
| | Acc | R | F1 | Acc | R | F1 |
| α=0.1 | 0.345 | 0.117 | 0.099 | 0.582 | 0.535 | 0.547 |
| α=0.3 | 0.337 | 0.110 | 0.096 | 0.551 | 0.497 | 0.506 |
| α=0.5 | 0.345 | 0.118 | 0.103 | 0.549 | 0.495 | 0.504 |
| α=0.7 | 0.344 | 0.116 | 0.099 | 0.571 | 0.522 | 0.534 |
| α=0.9 | 0.356 | 0.126 | 0.111 | 0.571 | 0.522 | 0.633 |
| **ADHCN(our)** | **0.646** | **0.443** | **0.487** | **0.706** | **0.650** | **0.647** |

| Model | Dblp | | | PatentDB | | |
|---|---|---|---|---|---|---|
| | Acc | R | F1 | Acc | R | F1 |
| α=0.1 | 0.715 | 0.698 | 0.702 | 0.837 | 0.508 | 0.496 |
| α=0.3 | 0.720 | 0.704 | 0.708 | 0.845 | 0.508 | 0.490 |
| α=0.5 | 0.722 | 0.706 | 0.711 | 0.851 | 0.507 | 0.486 |
| α=0.7 | 0.722 | 0.707 | 0.711 | 0.849 | 0.507 | 0.486 |
| α=0.9 | 0.719 | 0.704 | 0.707 | 0.845 | 0.506 | 0.487 |
| **ADHCN(our)** | **0.873** | **0.862** | **0.867** | **0.852** | **0.513** | **0.503** |

We adopted the adaptive representation fusion performed better than the model with the fusion method of weighted summation. Consequently, the adaptive representation fusion module can adaptively adjust the fusion of the two channels to obtain better model performance.

### 4.4 Different Adaptive Fusion

In this paper, we introduce the attention mechanism to achieve adaptive fusion. The process of fusion is shown in Eq.7 and Eq.8. Eq.7 takes the output $Z_l$ of the LE convolution channel, the output $Z_h$ of hypergraph convolution channel and the average representation $Z_c$ between them as the input of the attention mechanism to obtain the corresponding weights. Eq.8 sums the dot products of the obtained weights and their corresponding representations. In order to prove the effectiveness of this method, we set up experiments to compare our model with two methods. One is the adaptive fusion method directly using $Z_l$ and $Z_h$ without $Z_c$, the other is the adaptive fusion model proposed in [24], which the difference from ADHCN is the calculation method of $Z_c$. The Eq.11 defines the calculation method of $Z_c$:

$$Z_c = \frac{commConv(Z_h) + commConv(Z_h)}{2} \quad (11)$$

where $commConv(x)$ is the function defined by $W_c * x + b_c$. The $W_c$ and $b_c$ are the parameters to be trained. The experimental results are shown in Figure 3. In the figure, "ADHCN-Comm" represents the first comparison method, and "ADHCN+ComConv" represents the second comparison method.

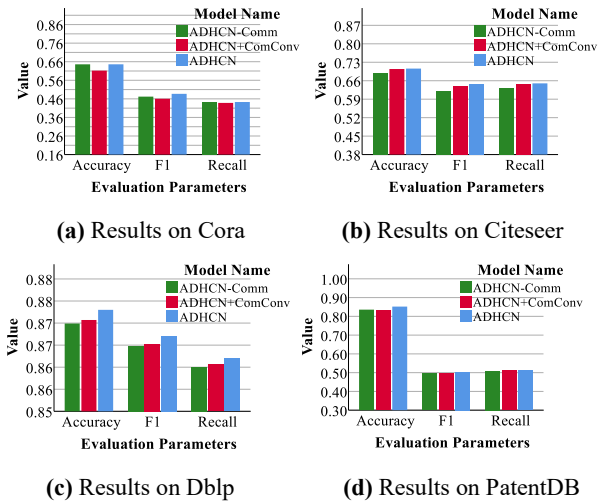

(a) Results on Cora  (b) Results on Citeseer

(c) Results on Dblp  (d) Results on PatentDB

**Figure 3** Comparison results of different adaptive fusion

The data in Figure 3 shows that the adaptive fusion method in ADHCN can achieve better performance on each dataset than those of the other two methods. Explicitly, the experimental results are most obvious on the Dblp dataset, where accuracy, F1 and recall are 0.2%, 0.1% and 0.2% higher than the second best, respectively. The method only fuses the outputs of the two convolution channels, which will lose the common information of the two channels. Another method using common convolution can avoid the loss of public information, but it increases the complexity of the model, thus resulting in poor model effect. ADHCN simplifies the model by replacing the common convolution part with the average of the two outputs, thus achieving better experimental results on all datasets.

## 5 Conclusions

In order to make full use of the high-order information in the technological intellectual property data, we used hypergraph to model the technological intellectual property entities and their relationships, and proposed an adaptive dual channel hypergraph convolution model to learn the constructed hypergraph, in which dual channels are used to prevent information loss caused by hypergraph expansion, and an attention mechanism is introduced to realize adaptive fusion between the two channels. Experimental results demonstrate that the suggested approach outperforms other comparable algorithms, and the proposed adaptive fusion mechanism improves the performance of the model.

## Acknowledgements

This work was supported by the National Natural Science Foundation of China (No.62192784, No.62172056).